\documentclass[aps,prb,preprint]{revtex4-1}

\usepackage{graphicx}
\begin{document}
%
%
\title{Morphology and magnetism of multifunctional nanostructured $\gamma$-Fe$_2$O$_3$ films: Simulation and experiments\\ \today}
\author{R. F. Neumann and M. Bahiana}
\affiliation{Instituto de F\'isica, Universidade Federal do Rio de Janeiro, Caixa Postal
68528, Rio de Janeiro 21941-972, Brazil}
\author{L.G. Paterno}
\affiliation{Universidade de Bras\'{\i}lia, Instituto de Qu\'{\i}mica, Bras\'{\i}lia DF 70910-900, Brazil}
%
\author{M.A.G. Soler}
\affiliation{Universidade de Bras\'{\i}lia, Instituto de F\'{\i}sica, Bras\'{\i}lia DF 70910-900, Brazil}
\author{J.P. Sinnecker}
\affiliation{Centro Brasileiro de Pesquisas F\'{\i}sicas, Rio de Janeiro RJ 22290-180, Brazil}
\author{J.G. Wen}
\affiliation{Electron Microscopy Center, Materials Science Division, Argonne National Laboratory, Argonne, Illinois 60439, United States}
\author{P.C. Morais}
\affiliation{Universidade de Bras\'{\i}lia, Instituto de F\'{\i}sica, Bras\'{\i}lia DF 70910-900, Brazil}
\begin{abstract}
This paper introduces a new approach for simulating magnetic properties of nanocomposites comprising magnetic particles embedded in a non-magnetic matrix, taking into account the 3D structure of the system in which particles' positions correctly mimic real samples. The proposed approach develops a multistage simulation procedure in which the size and distribution of particles within the hosting matrix is firstly attained by means of the Cell Dynamic System (CDS) model. The 3D structure provided by the CDS step is further employed in a Monte Carlo (MC) simulation of zero-field-cooled/field-cooled (ZFC/FC) and magnetic hysteresis loops ($M \times H$ curves) for the system. Simulations are aimed to draw a realistic picture of the as-produced ultra-thin films comprising maghemite nanoparticles dispersed in polyaniline. Comparison (ZFC/FC and $M \times H$ curves) between experiments and simulations regarding the maximum of the ZFC curve ($T_{\mbox{\scriptsize MAX}}$), remanence ($M_R/M_s$) and coercivity ($H_C$) revealed the accuracy of the multistage approach proposed here while providing information about the system's morphology and magnetic properties. For a typical sample the value we found experimentally for $T_{\mbox{\scriptsize MAX}}$ (54 K) was very close to the value provided by the simulation (53 K). For the parameters depending on the nanoparticle clustering the experimental values were consistently lower ($M_R/M_s$ = 0.32 and $H_C$ = 210 Oe) than the values we found in the simulation ($M_R/M_s$ = 0.53 and $H_C$ = 274 Oe). Indeed, the approach introduced here is very promising for the design of real magnetic nanocomposite samples with optimized features.
\end{abstract}

\maketitle


%
%
\section{Introduction}
Assemblies of nanoscaled particles or clustered nanosized systems have attracted a great deal of interest either due to their fundamental properties or their potential use in several technological applications \cite{fan2010self,grzelczak2010directed}. In particular, composites containing nanosized magnetic particles dispersed in a hosting matrix can be envisaged as multifunctional materials with promising applications in emerging fields \cite{balazs2006nanoparticle,kim2002multifunctional,nunes2010multifunctional}. In this particular class of material system isolated magnetic particles are sufficiently small, so that they usually remain in a single-domain magnetic state \cite{bedanta2009supermagnetism}. Nonetheless, it is possible to modulate the particle-particle interaction over a broad range of magnetic particle concentration in such a way that inter-particle interaction starts playing a significant role on the nanocomposite end properties \cite{papaefthymiou2009nanoparticle,rebolledo2008signatures}. Different types of magnetic interactions are found in nanoparticulated systems, including the magnetic dipole coupling; the indirect exchange or Rudermann-Kittel-Kasuya-Yosida (RKKY) interaction mediated by conduction electrons, for instance when the matrix and particles are both metallic; the short-range direct exchange interaction when surfaces of neighbor nanoparticles are in close contact; and the tunneling exchange interaction, which occurs when nanoparticles are considered to be only some tenths of nanometers apart from one another \cite{jensen2003low,skomski2008simple}.

Besides the complexity of the dipolar interaction there is also an uncertainty in determining the anisotropy axis direction and location of the magnetic nanoparticle within the nanocomposite matrix. However, advances in theoretical, experimental and simulation have contributed to unveil the magnetic behavior of one and two dimensional systems of interacting particles \cite{chantrell1996modelling,bahiana2004ordering,bastos2002role,dimitrov1996magnetic,dormann1998effect,garcia2000influence,iglesias2004magnetic,kechrakos2005monte,weis2003simulation}. Monte Carlo (MC) simulations have been used to assess the magnetic properties of nanocomposites comprising iron oxide-based nanosized particles embedded in non-magnetic matrices. Buj\'an-N\'u\~nez et al. \cite{bujan2008influence} studied the influence of the nanoparticle size on the blocking temperature of interacting systems (dipole–dipole interaction) using MC simulation of zero-field-cooled curves. Hoppe et al. \cite{hoppe2008effect} employed a similar method to study the influence of the dipolar interaction on the blocking temperature of a random array of superparamagnetic particles (maghemite) dispersed within different polymeric matrices. Lamba and Annapporni \cite{lamba2004single} reported simulation results including effects of disorder, short-range exchange, long-range dipolar interaction and anisotropy effects, using parameters of a maghemite-based system.

It becomes clear that the simulation of the magnetic response of realistic 3D systems of interacting nanoparticles is mandatory to understand the relationship between the nanocomposite structure and its magnetic properties. Despite efforts expended to address this issue not many investigations have been developed with great level of detail. For instance, assemblies of monodisperse particles placed at lattice positions have been used to describe experimental systems with much more complex morphology. Another possible approach, but rather inefficient, sorts particle sizes from a distribution and places them at random positions. Since particles cannot overlap, the rejection frequency for new positions gets higher while the sample becomes denser and, consequently, the algorithm becomes very slowly.

In the present study we propose an alternative approach, a multistage simulation in which the 3D spatial structure of a nanocomposite system comprising magnetic nanoparticles dispersed within a non-magnetic matrix is firstly generated by means of the Cell Dynamic System (CDS) model \cite{oono1988study,puri1988study}. With the use of an iterative algorithm the model is capable of generating the main systems' features, namely particle positions, sizes, and shapes. In a second step, the generated 3D structure is employed in the MC simulation of zero-field-cooled/field-cooled (ZFC/FC) curves and hysteresis loops. The simulated 3D structure and the corresponding magnetic properties are then compared with the experimental data recorded from real samples, namely ultra-thin films of polyaniline (PANI) encapsulating citrate-coated maghemite (cit-MAG) nanoparticles.
\section{Experimental Procedure}
\subsection{Synthesis of Nanoparticle dispersions and Preparation of Nanostructured Films}
Magnetic nanocomposites were produced via the layer-by-layer (LbL) technique following the procedure previously reported \cite{paterno2010magnetic,paterno2009layer,soler2012assembly}, in which layers of anionic citrate-coated maghemite (cit-MAG) nanoparticle and cationic polyaniline (PANI) were alternately and electrostatically adsorbed onto a solid substrate. The cit-MAG suspension was prepared in a two-step chemical route as reported elsewhere \cite{morais2006preparation} and briefly described as follows. In the first step Fe(II) and Fe(III) ions were co-precipitated in alkaline medium to produce magnetite nanoparticle, which was oxidized to produce maghemite. In the second step, the as-produced maghemite nanoparticles were rendered negatively charged via functionalization with excess of citric acid and suspended as an aqueous-based magnetic fluid sample. The doped PANI solution was prepared by dissolution of a commercial undoped PANI (Mw 10,000 g mol$^{-1}$, Aldrich, USA) in a mixture of N',N''-dimethylacetamide and hydrochloridric aqueous solution (pH 2.7). The nanocomposite production was realized by alternate and successive immersions of hydrophilic (100)-oriented Si stripes into PANI (polycation) and cit-MAG (anion) solutions. In between immersions substrates were rinsed in a stirred bath of HCl solution (pH 2.7) and dried with N$_2$ flow. Multilayered magnetic nanocomposite films of (PANI/cit-MAG)$_{\ell}$, with $\ell$ increasing from 1 to 50 bilayers, were produced by repeating the cycles described above. Samples with $\ell = 10, 25$ and 50 were systematically investigated. All LbL depositions were carried out at room temperature ($\approx 25\;^\circ$C).
\subsection{Measurements and Instruments}
Surface morphology of the thin composite films was characterized with scanning electron microscope (SEM) (Hitachi S4800). Transmission electron microscopy (TEM) micrographs were recorded with a 200 kV instrument (JEOL JEM 2100) to evaluate the average particle diameter ($D_{\mbox{\scriptsize TEM}} = (7.5 \pm 0.1)$ nm) of nanoparticles in the magnetic fluid sample as well as their distribution within the nanocomposite films (cross-section images).
  
Preparation of the Si substrate-deposited nanocomposite films for cross-sectional TEM observation followed the standard protocol: films were firstly glue face-to-face, then Si substrate was mechanically ground down to about 20 $\mu$m in thickness, following the ion milling of the sample to perforation.

Magnetic properties of the nanocomposite samples, at different applied fields and temperatures, were measured using a Cryogenics S600 SQUID magnetometer. 
 The total magnetization of each nanocomposite sample was scaled to the $\%$wt content of cit-MAG, the latter obtained from UV-vis spectroscopy (Shimadzu spectrophotometer, model UVPC 1600). The $\%$wt contents of cit-MAG within the nanocomposite samples were obtained by measuring their optical absorption at 480 nm wavelength. The cit-MAG contents were determined from a previously constructed calibration curve based on the absorbance values of cit-MAG suspensions, at different nanoparticle concentrations and at 480 nm, which corresponds to the typical wavelength for electronic transitions in iron oxides \cite{sherman1985electronic}.
\section{Simulation}
The simulation procedure comprises three main stages: generation of the spatial structure, attribution of the magnetic properties and simulation of magnetization curves. For each part different computational techniques were used, as described below.
\subsection{Three dimensional simulation of the sample's morphology}
The (PANI/cit-MAG)$_\ell$ nanofilms investigated are a typical binary system with nanosized magnetic particles embedded in a polymeric matrix. The PANI phase surrounding cit-MAG nanoparticles prevents particle clustering leading to a structure with some regularity in terms of particle shape and position. In this binary material system, the sample can be idealized as an array of isolated spheres, with a well-defined size distribution, packed in a local HCP order and presenting long-range disorder. The pattern observed in the film cross-sectional micrograph resembles those in samples consisting of microphase separated diblock copolymers, in which one of the polymer subchains is much longer than the other one. This similarity motivated us to select the CDS model for the phase segregation of block copolymers to simulate the morphology of the as-produced nanostructured magnetic films. The CDS technique was originally proposed by Oono and Puri \cite{oono1988study,puri1988study} and has been successfully employed to simulate the spatial structure of block copolymers \cite{bahiana1990cell} with great computational efficiency. Let us remind that the polymeric matrix itself surrounding the maghemite nanoparticles we study in this paper has no relation with the diblock copolymers that served as an inpiration for the use of the CDS technique. It is only the final pattern generated by the CDS numerical algorithm that resembles the (PANI/cit-MAG)$_\ell$ nanofilms, thus motivating us to use this specific technique.

The CDS technique proposes models discrete in time and space. It is based on the sequential iteration of diffusely coupled maps that mimic the local segregation dynamics in terms of a local order parameter, $\psi(\vec r,t)$, providing the number density difference between the two segregating species at the position $\vec r$ and time $t$. Non-segregated regions correspond to cells in which $\psi \approx 0$. Segregated domains have either $\psi > 0$ or $\psi < 0$, depending upon the predominant species. In the present case we have nanoparticles with fixed sizes, so we must seek a model in which the final segregated pattern has the same characteristic. Although the model provides the time evolution of the segregation process, we are interested only in the final pattern consisting of roughly spherical particles placed in a matrix with some local order. The iteration procedure converges to this pattern very quickly, so the method is extremely efficient. 

In the present study we consider a three dimensional lattice with periodic boundary conditions along all directions. The order parameter was defined such that he final pattern had spherical domains with $\psi < 0$ (maghemite nanoparticles) in a background with $\psi > 0$ (polymeric matrix). The system was initialized with  $\psi(\vec r, t) = \psi_0 + \alpha(\vec r)$, where $\psi_{\mbox{\tiny 0}}$ is the off-criticality and $\alpha$ is a random number in the interval [-0.005,0.005]. The order parameter at time $t+1$ is computed from the previous one according to Eq. \ref{Eq_CDS1}:
\begin{equation}
  \psi (\vec r,t+1)  =  \psi (\vec r,t) - \nabla^2 \left[I(\vec r,t)\right] +  B \left[ \psi_{\mbox{\tiny 0}} - \psi (\vec r,t) \right],
  \label{Eq_CDS1}
\end{equation}
where
\begin{equation}
  I(\vec r,t)  = A \tanh \left[\psi (\vec r,t)\right] - \psi (\vec r,t) + \Gamma \nabla^2 \left[\psi (\vec r,t)\right].
  \label{Eq_CDS2}
\end{equation}
Here, $A > 1$ is the segregation strength and defines the maximum absolute value of $\psi$. $\Gamma$ is a phenomenological diffusion constant and the parameter $B \ll 1$ is responsible for the partial segregation. The symbol $\nabla^2$ stands for the three-dimensional isotropic discrete Laplacian defined as \cite{shinozaki1993spinodal}:
\begin{equation}
 \nabla^2 [f(\vec r)]  = \left[ \frac{6}{80} \sum_{\mbox{\scriptsize nn}} f(\vec r_{\mbox{\tiny nn}}) + \frac{3}{80} \sum_{\mbox{\scriptsize nnn}} f(\vec r_{\mbox{\tiny nnn}}) \right.
+ \left. \frac{1}{80} \sum_{\mbox{\scriptsize nnnn}} f(\vec r_{\mbox{\tiny nnnn}}) \right] - f(\vec r) ,
\label{Eq_laplacian}
\end{equation}
where the subscripts nn, nnn and nnnn stand for ``nearest-neighbor'', ``next-nearest-neighbor'' and ``next-next-nearest-neighbor'', respectively. 

The spatial structure to be analyzed was simulated by iterating 2000 times Eqs. \ref{Eq_CDS1} and \ref{Eq_CDS2} on a $300 \times 300 \times 30$ slab, with the $300 \times 300$ planes parallel to the $xy$ plane. In all simulations we used the parameter values that are know to generate the pattern we are looking for \cite{bahiana1990cell}: $\psi_0 = 0.3, A = 1.3, \Gamma = 0.5$, and $B = 0.02$. 
\subsection{Analysis of the simulated morphology}
After the CDS simulation one has the values of order parameter for all cells. The next step is to convert this information into $N_p$ sets of parameters $\{\vec r_i, \vec m_i, K^s_i, \hat e^s_i\}$ defined as position ($\vec r_i$), total magnetic moment ($\vec m_i$), shape anisotropy constant ($K^s_i$) and easy magnetization axis ($\hat e^s_i$) of the $i$-th  nanoparticle.

The first step is to identify the particles that are associated to the clusters with $\psi<0$. For that, a three-dimensional extension of the Hoshen \& Kopelman algorithm \cite{hoshen1976percolation} for cluster identification and labeling was used. This analysis lead to the total number of clusters ($N_p$) and to the number of cells belonging to each cluster ($n_i$). Afterwards, we determined the clusters' positions  by calculating the center of mass of the group of cells forming each cluster. Next comes the analysis of the particles´ shape for determination of shape anisotropy properties (easy axis direction and strength) of each cluster. We considered that the particles had a prolate ellipsoidal shape. In this case the shape anisotropy constant could be calculated as follows \cite{cullity2011introduction}:
\begin{equation}
 K^s = \left( \pi - \frac{3}{4} N_c \right) M_s^2\; ,
 \label{Eq_Ks}
\end{equation}
with
\begin{equation}
 N_c(\rho=c/a) = \frac{4 \pi}{\left( \rho^2-1 \right)} \left[ \frac{\rho}{\sqrt{\rho^2-1}} \ln \left(\rho + \sqrt{\rho^2-1} \right) - 1 \right]\; ,
 \label{Eq_demag_factor}
\end{equation}
where $a$ and $c$ are lengths of the shorter and longer particle axis, respectively. 
The ratio $\rho = c/a$ and the easy axis direction can be determined by diagonalizing the inertia matrix for each particle. 

Another issue is the calibration of the length scale since the CDS model has arbitrary units. In order to obtain a good quantitative comparison with experiment we calibrated the length scale of the simulated pattern by defining that the simulated average diameter ($D_{\mbox{\scriptsize CDS}}$) had the same value as in the real sample ($D_{\mbox{\scriptsize TEM}}$), that is $D_{\mbox{\scriptsize CDS}} \cong 7.5$ nm. With this, the length of each cell is
\begin{equation}
 d_{\mbox{\scriptsize cell}} = D_{\mbox{\scriptsize TEM}} \left( \frac{\pi}{6 \bar{n}_{\mbox{\scriptsize c}}} \right)^{\frac{1}{3}}\; ,
 \label{Eq_l_cell}
\end{equation}
where $\bar{n}_{\mbox{\scriptsize c}}$ is the average number of cells in the clusters.
The magnetic moment of each cluster depends on its number of cells ($n_i$) and can be obtained from the experimental value of the saturation magnetization ($M_s$) through $|\vec{m}_i| = M_s n_i d_{\mbox{\scriptsize cell}}^3$. For this simulated system, the average shape anisotropy constant was $\langle K^s\rangle = 7.2\times 10^4$ erg/cm$^3$ as calculated through Eq. \ref{Eq_Ks}.
\subsection{Monte Carlo simulation}
In order to proceed with the simulation of the magnetization curves it is necessary to specify which energy terms will be taken into account in the MC simulation. According to our simulated ensemble, the distance $R_{\mbox{\scriptsize nn}}$ between nearest neighboring grains was relatively large, so that direct exchange and RKKY interactions were neglected. Within this approximation, the magnetic dipolar interaction was the only one considered. We wrote down this interaction in terms of magnetic dipoles positioned at the center of the particles. Including anisotropy and Zeeman terms, we defined the total energy of the $N_p$ particles in the presence of an external magnetic field ($\vec H$) as:
\begin{equation}
E  =  \sum_{i=1}^{N_p} \left[
 -\vec m_i \cdot \vec H 
 + E^{\mbox{\tiny A}}_i   
 + \sum_{j > i}^{N_p} \frac{\vec m_i \cdot \vec m_j - 3(\vec m_i \cdot \hat r_{ij})(\vec m_j \cdot \hat r_{ij})}{ |\vec r_{ij}|^3} 
\right] ,
 \label{Eq_energy}
\end{equation}
where the first term denotes the particle's Zeeman term and $E^{\mbox{\tiny A}}_i$ accounts for the anisotropy energy which includes shape and crystalline contributions. Based on the experimental data on maghemite available from the literature \cite{dutta2004magnetic,hendriksen1994ultrafine,hergt2004maghemite,jonsson1995aging,martinez1998low,martinez1996magnetic,shendruk2007effect,vassiliou1993magnetic}
 We considered uniaxial crystalline anisotropy, using the value $K^c = 1.25 \times 10^5$ erg/cm$^3$, and random cristalline orientation ($\hat e^c_i$) for each particle. Whith this, 
\begin{equation}
E^{\mbox{\tiny A}}_i = -K^s_i V_i \left( \frac{\vec m_i \cdot \hat e^s_i}{|\vec m_i|} \right)^2 
- K^c V_i \left( \frac{\vec m_i \cdot \hat e^c_i}{|\vec m_i|} \right)^2\; .
\label{Eq_anisotropy}
\end{equation}
The third term in Eq. \ref{Eq_energy} refers to the particle-particle dipolar interaction, in which $\vec{r}_{ij}$ is the distance between particles $i$ and $j$. In order to mimic the nanocomposite film we have applied periodic boundary conditions for the magnetostatic interaction on the $xy$ plane.

The MC simulations were performed using the Metropolis algorithm. In short, a new orientation for the magnetic moment of a randomly chosen nanoparticle is set, then the energy difference ($\Delta E$) between old and new orientations is calculated and accepted with probability $p = \min \left[ 1, \exp (-\Delta E / k_{\mbox{\scriptsize B}} T)  \right]$. The new orientation for the magnetic moments is chosen according to an algorithm analogous to the one proposed by Nowak et al. \cite{nowak2000monte}, in which a randomly oriented vector $\vec v$ with modulus $G|\vec{m}_i|$ is added to the moment $\vec{m}_i$. By changing $G$ one can control how different new and old directions might be. With this algorithm the magnetic moment is restricted to a cone centered in $\vec{m}_i$ with an opening angle $\theta_{\mbox{\scriptsize max}} = \sin^{-1}(G)$ (see Fig. \ref{Fig_vectors}). We have used $\theta_{\mbox{\scriptsize max}} = 15^\circ$ throughout the simulation. One MC step (MCS) corresponds to examining, in average, all the magnetic moments.
\section{Results and Discussion}
The (PANI/cit-MAG)$_ell$  nanofilm is a typical binary system composed by nanosized magnetic particles embedded within the polymeric matrix, as observed in the cross-sectional micrograph of a (PANI/cit-MAG)10 sample containing ten (n = 10) bilayers (Figure 1a). The PANI phase surrounds cit-MAG nanoparticles and prevents clustering to some extent. Clustering in magnetic nanoparticles systems is an undesirable effect as it introduces strong dipolar interactions. Among other effects, the close proximity of nanoparticles may induce the transition from superparamagnetic to ferro(ferri)magnetic states which, therefore, place huge difficulties on the understanding of their magnetic properties.
A qualitative analysis shows good agreement between both structures.
This binary material system can be idealized by an array of individual and isolated spheres, with a well-defined average radius and size distribution, packed in a local HCP order but presenting long-range disorder. The real sample is prepared from a colloidal suspension composed by monodomain nanosized particles (cit-MAG), whith average diameter $D_{\mbox{\scriptsize TEM}} = 7.5$ nm determined by transmission electron microscopy (TEM not shown). 
This information was used to determine the length scale of the simulated ensemble.  Additionally, the pair distribution function, $g(r)$, attained by the CDS simulation (Figure \ref{Fig_g_r_}) clearly indicates the nanoparticles' ensemble has a short-range order at distances shorter than 30 nm. From the first $g(r)$ peak position (see Figure \ref{Fig_g_r_}(b)) one determines the average center-to-center nearest neighbor distance (Rnn) equal to 10 nm. The correspondent experimental value determined from HRTEM images as observed in the inset of Figure \ref{Fig_g_r_}(a) is also 10 nm. It appears that morphological features simulated by the CDS model correlate quite well to those presented by the real sample after HRTEM micrographs. 

Considering that the average particle diameter is 7.5 nm, the average interparticle separation is about 2.5 nm. In the real sample this number represents the average distance of neighboring particles, that is a surface-to-surface spacing of about 2.5 nm, which is filled by the hosting polymeric material, in very good agreement with the observed TEM cross-sectional micrographs in Fig. \ref{Fig_g_r_}(a).

The simulated ZFC/FC curves were obtained starting from a demagnetized state at 2 K. The temperature was then increased at a rate of $\nu = 50$ MCS/2K, up to 300 K, under an applied in-plane magnetic field of 36 Oe. Here MCS stands for Monte Carlo steps. For the simulation of the FC curve we started the simulation  with the last configuration of the ZFC curve and the temperature was decreased down to 2 K at the same rate as in the ZFC curve. This procedure was repeated 2000 times and the results were averaged out over all statistically equivalent realizations. For simplification we have considered $K^s_i = \langle K^s\rangle$ (its average over the ensemble). Experimental and simulated ZFC/FC curves are presented in Figures \ref{Fig_ZFC-FC}(a) and \ref{Fig_ZFC-FC}(b), respectively. The values of $T_{\mbox{\scriptsize MAX}}$ in the ZFC curves were 54 and 53 K for the experimental and simulated curves, respectively. Since $T_{\mbox{\scriptsize MAX}}$  scales with the blocking temperature $T_{\mbox{\scriptsize B}}$ we can draw conclusions about the amount of interaction in the samples through the variation of  $T_{\mbox{\scriptsize MAX}}$. The decreasing of $\ell$ (the number of bilayers) in the (PANI/cit-MAG)$_\ell$ magnetic nanocomposite films,  causes the reduction of $T_{\mbox{\scriptsize MAX}}$ from 54 K ($\ell = 50, 25$) to 45 K ($\ell = 10$). This is consistent with the picture of particle interaction effects reported in the literature \cite{papusoi1999particle} as the content of the magnetic nanoparticle incorporated within (PANI/cit-MAG)$_\ell$ reduces with the reduction of $\ell$. The small discrepany in $T_{\mbox{\scriptsize MAX}}$ from 53 K (simulated) in comparison to 54 K (experimental), as shown in Figures \ref{Fig_ZFC-FC}, indicates that the two samples (simulated and experimental) are indeed quite similar, though the real sample might present stronger particle-particle interaction than the simulated one. Interesting to note that $T_{\mbox{\scriptsize MAX}} = 54$ K was observed in both nanocomposite films ($\ell = 50, 25$), indicating that the nanoparticle packing has reached a limit and the only difference between the two samples is actually the film thickness. Furthermore, while comparing with the simulated ZFC/FC curves the observed flattening of the experimental ZFC/FC curves ($\ell = 10, 25, 50$) indicates the mean-field effect, more likely due to the increasing of the sample's demagnetizing factor as a consequence of the particle chain formation \cite{papusoi1999particle}. While the simulated ZFC/FC curves shown in Fig. \ref{Fig_ZFC-FC}(b) were derived from a system comprising an ensemble of single nanoparticles, though including dipolar interaction among them, as described in Eq. \ref{Eq_energy}, the experimental data do indicate that particle clustering takes place in the real samples we investigated here.

While comparing experimental and simulated ZFC/FC curves displayed in Fig. \ref{Fig_ZFC-FC}, one notes that the FC curve (Fig. \ref{Fig_ZFC-FC}(a)) obtained for the real sample is rather flat below the blocking temperature, whereas the one obtained by Monte Carlo simulations presents a maximum peak in the same temperature range.
 As mentioned earlier, the dipolar interaction acting in the real system would be greater than that it is being considered in simulations. Nonetheless, the CDS and MC combined were still capable of detecting, even though qualitatively, the influence of the film densification on its end properties. For example, we have succeeded in controlling the average particle-particle distance within the nanofilms by simply varying the number of deposited bilayers or else by changing the nanoparticle's concentration in the colloidal dispersion used for film depositions. In parallel, with the CDS model we have simulated hypothetical ultra-thin slabs of three different thicknesses with a fixed number of nanoparticles ($N_p\approx 3000$). As to real samples, the blocking temperature displayed a monotonic increase as the film became denser either by depositing more bilayers or by increasing the concentration of the colloidal dispersion. In addition, effective barrier energy demanded for magnetic moment reversal also increased as the film became denser, as determined by ac susceptibility measurements performed on the same real samples. This effect is ascribed to the decrease in particle-particle which consequently turns stronger the dipolar interaction.\cite{papusoi1999particle} From magnetic ZFC/FC curves obtained after MC simulation of the hypothetical slabs it was observed a similar trend with an increase of blocking temperatures as the slabs became thinner (or denser). These results allow one to conclude that despite some failure, the combination of CDS model and MC simulation is able to predict in a rather qualitative way the morphology and magnetic properties of layer-by-layer assembled magnetic nanocomposites. Moreover, further adjustments of the approach may allow for a more detailed description of magnetic nanocomposites in general.

The discussion regarding the data provided by the ZFC/FC curves is corroborated by the experimental ($\ell = 25$) and simulated hysteresis curves shown in Figures \ref{Fig_MxH}(a) and \ref{Fig_MxH}(b), respectively. 

\section{Conclusions}
A computational procedure was developed to predict the magnetic properties of a nanocomposite material of simulated morphologies. The approach combines two simulation procedures. The first one, based on the Cell Dynamic System model, is capable of generating hypothetical pieces of a solid with nanoparticles dispersed in a continuous matrix and sorted at pre-defined inter-particle distances. The distribution function, g(r), attained by the CDS simulation when considering nanoparticles of 7.5 nm in diameter provided the center-to-center distance between two nanoparticles equal to 10 nm and, consequently, a surface-to-surface distance of about 2.5 nm. The simulated value was in good agreement with the nanoparticle's surface-to-surface distance of about 3 nm determined from HRTEM micrographs of real maghemite/polyaniline layer-by-layer films. The second step of the computational procedure applies Monte Carlo simulation to predict magnetic properties of the as-simulated solid morphology. The MC simulates both ZFC/FC and hysteresis curves after determining the energy of magnetic nanoparticles under influence of the magnetic field, while taking into account dipolar interactions between particles as well as anisotropic energies such as those due to shape and crystalline structure. In many instances, the simulated magnetic properties correlated quite well with the properties experimentally measured. On the other hand, the simulated curves could not detect the strong dipolar interaction placed between particles below the blocking temperature. This failure was attributed to the fact that the CDS simulation did not account for the presence of nanoparticles's aggregates which are inherent to the adsorption of nanoparticles onto the substrate. However, this particular feature could be included in order to improve the simulation. 
In summary, the multistage simulation developed herein has proved to be quite promising once it enables one to prepare magnetic nanocomposites with optimized morphology and predictable magnetic properties. The right adjustment of simulation parameters is still a challenge but perfectly attainable. This novel approach can find application in fields beyond that investigated here, spanning from biology and medicine to engineering and nanoelectronics, where phase segregated systems play a major role.

%
%
%
%
%
\section{Acknowledgments}
R. F. Neumann and M. Bahiana acknowledge support from CNPq and FAPERJ. The financial support from the Brazilian agencies MCT-CNPq, FINEP, CAPES, is also gratefully acknowledged. M. A. G. Soler thanks Professor Steve Granick (Department of Materials Science and Engineering, University of Illinois at Urbana-Champaign, USA) for the hospitality in the period from April to June, 2009, and CAPES-Brazil (4410-08-4). The authors acknowledge the support of Dr. Michael Marshall (Frederick Seitz Materials Research Laboratory, University of Illinois at Urbana-Champaign, USA) in the TEM measurements, Dr. Emilia C.D. Lima (Universidade Federal de Goiás, Brazil) for supplying colloidal samples and Prof. Miguel A. Novak (Universidade Federal do Rio de Janeiro, Brazil) for magnetic characterization facilities. 
%

%
%
%
\pagebreak
\begin{figure}
\includegraphics[width=15cm]{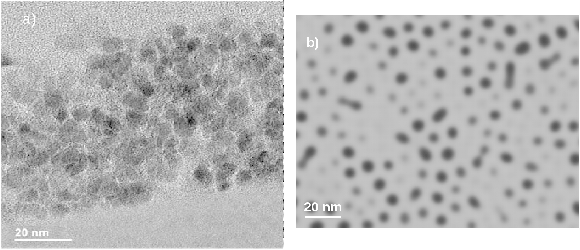}
\caption{Typical surface morphology of a (PANI/cit-MAG)$_{10}$ film (a) and simulated (b) cross-section structure after CDS modeling.}
\label{Fig_morphology}
\end{figure}

\begin{figure}
\includegraphics[width=7cm]{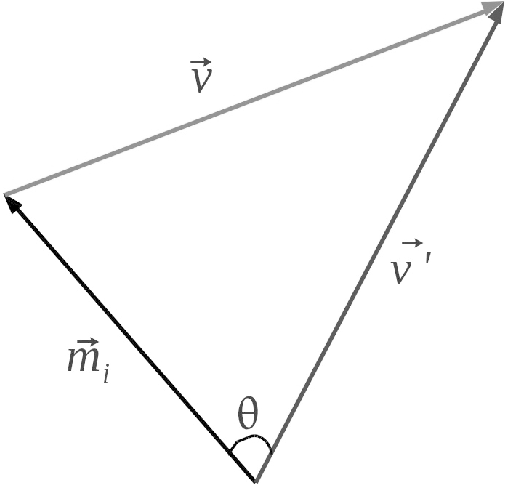}
\caption{At each trial step a magnetic moment $\vec{m}_i$ (black) is selected. A random vector $\vec v$ with modulus $G|\vec{m}_i|$ (red) is added to the latter. The resulting vector $\vec v'$ (blue) gives the direction of the new magnetic moment. After rescaling it in order to keep the same modulus as $\vec{m}_i$, it will be the new magnetic moment $\vec{m'}_i$ We can control the angle $\theta$ between $\vec{m}_i$ and $\vec{m'}_i$ by choosing the appropriate $G$.}
\label{Fig_vectors}
\end{figure}
\begin{figure}
\includegraphics[width=10cm]{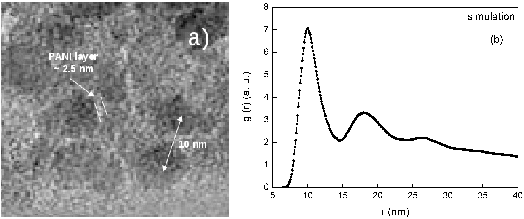}
\caption{Typical cross-section TEM micrograph of a (PANI/cit-MAG)$_{10}$ film (a); and simulated pair distribution function, $g(r)$, for the $300\times 300\times 30$ grid (b).}
\label{Fig_g_r_}
\end{figure}
\begin{figure}
\includegraphics[width=15cm]{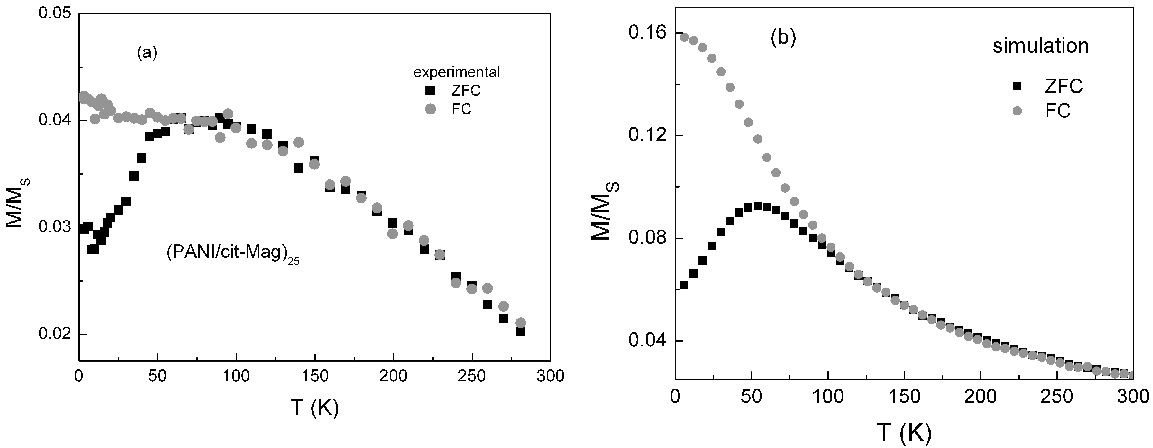}
\caption{Experimental (PANI/cit-MAG)$_{25}$ nanofilm (a) and simulated (b) ZFC/FC curves obtained at a steady field of 36 Oe. Solid lines on (b) are guide to the eyes only.}
\label{Fig_ZFC-FC}
\end{figure}
\begin{figure}
\includegraphics[width=15cm]{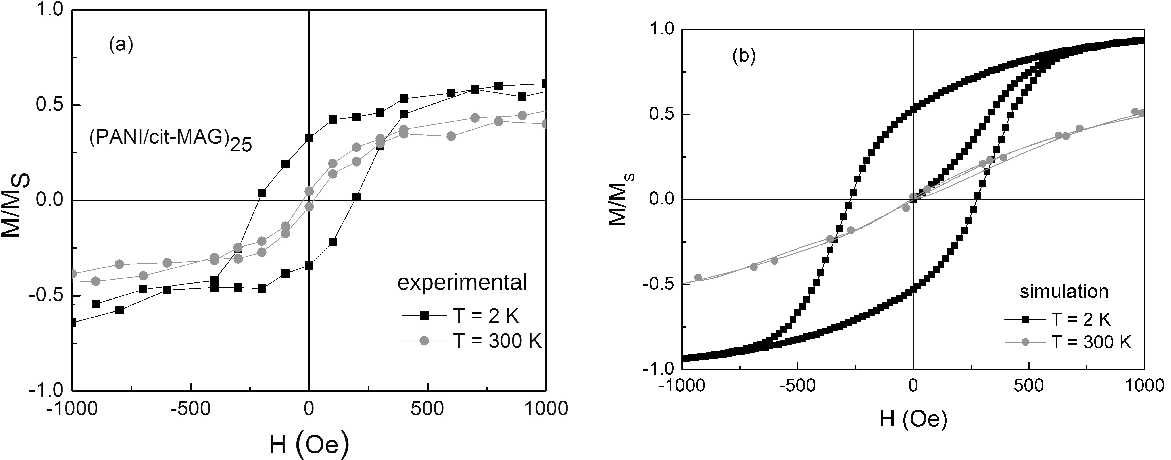}
\caption{Experimental (PANI/cit-MAG)$_{25}$ nanofilm (a) and simulated (b) hysteresis curves at different temperatures. Solid lines on (b) are guide to the eyes only.}
\label{Fig_MxH}
\end{figure}


\begin{thebibliography}{10}

\bibitem{fan2010self}
J.A. Fan, C.~Wu, K.~Bao, J.~Bao, R.~Bardhan, N.J. Halas, V.N. Manoharan,
  P.~Nordlander, G.~Shvets, and F.~Capasso.
\newblock Self-assembled plasmonic nanoparticle clusters.
\newblock {\em Science}, 328(5982):1135--1138, 2010.

\bibitem{grzelczak2010directed}
M.~Grzelczak, J.~Vermant, E.M. Furst, and L.M. Liz-Marza?n.
\newblock Directed self-assembly of nanoparticles.
\newblock {\em ACS Nano}, 4(7):3591--3605, 2010.

\bibitem{balazs2006nanoparticle}
A.C. Balazs, T.~Emrick, and T.P. Russell.
\newblock Nanoparticle polymer composites: where two small worlds meet.
\newblock {\em Science}, 314(5802):1107--1110, 2006.

\bibitem{kim2002multifunctional}
H.S. Kim, B.H. Sohn, W.~Lee, J.K. Lee, S.J. Choi, and S.J. Kwon.
\newblock Multifunctional layer-by-layer self-assembly of conducting polymers
  and magnetic nanoparticles.
\newblock {\em Thin Solid Films}, 419(1):173--177, 2002.

\bibitem{nunes2010multifunctional}
J.~Nunes, K.P. Herlihy, L.~Mair, R.~Superfine, and J.M. DeSimone.
\newblock Multifunctional shape and size specific magneto-polymer composite
  particles.
\newblock {\em Nano Letters}, 10(4):1113--1119, 2010.

\bibitem{bedanta2009supermagnetism}
S.~Bedanta and W.~Kleemann.
\newblock Supermagnetism.
\newblock {\em Journal of Physics D: Applied Physics}, 42:013001, 2009.

\bibitem{papaefthymiou2009nanoparticle}
G.C. Papaefthymiou.
\newblock Nanoparticle magnetism.
\newblock {\em Nano Today}, 4(5):438--447, 2009.

\bibitem{rebolledo2008signatures}
A.F. Rebolledo, A.B. Fuertes, T.~Gonzalez-Carre{\~n}o, M.~Sevilla,
  T.~Valdes-Solis, and P.~Tartaj.
\newblock Signatures of clustering in superparamagnetic colloidal
  nanocomposites of an inorganic and hybrid nature.
\newblock {\em Small}, 4(2):254--261, 2008.

\bibitem{jensen2003low}
P.J. Jensen and G.M. Pastor.
\newblock Low-energy properties of two-dimensional magnetic nanostructures:
  interparticle interactions and disorder effects.
\newblock {\em New Journal of Physics}, 5:68, 2003.

\bibitem{skomski2008simple}
R.~Skomski.
\newblock {\em Simple models of magnetism}.
\newblock Oxford University Press, USA, 2008.

\bibitem{chantrell1996modelling}
R.W. Chantrell, G.N. Coverdale, M.~El-Hilo, and K.~O'Grady.
\newblock Modelling of interaction effects in fine particle systems.
\newblock {\em Journal of Magnetism and Magnetic Materials}, 157:250--255,
  1996.

\bibitem{bahiana2004ordering}
M.~Bahiana, J.P. Pereira~Nunes, D.~Altbir, P.~Vargas, and M.~Knobel.
\newblock Ordering effects of the dipolar interaction in lattices of small
  magnetic particles.
\newblock {\em Journal of Magnetism and Magnetic Materials}, 281:372--377,
  2004.

\bibitem{bastos2002role}
C.S.M. Bastos, M.~Bahiana, W.C. Nunes, M.A. Novak, D.~Altbir, P.~Vargas, and
  M.~Knobel.
\newblock Role of the alloy structure in the magnetic behavior of granular
  systems.
\newblock {\em Physical Review B}, 66(21):214407, 2002.

\bibitem{dimitrov1996magnetic}
D.A. Dimitrov and G.M. Wysin.
\newblock Magnetic properties of superparamagnetic particles by a monte carlo
  method.
\newblock {\em Physical Review B}, 54(13):9237, 1996.

\bibitem{dormann1998effect}
J.L. Dormann, L.~Spinu, E.~Tronc, J.P. Jolivet, F.~Lucari, F.~D'orazio, and
  D.~Fiorani.
\newblock Effect of interparticle interactions on the dynamical properties of
  $\gamma-\mbox{Fe}_2\mbox{O}_3$ nanoparticles.
\newblock {\em Journal of Magnetism and Magnetic Materials}, 183(3):L255--L260,
  1998.

\bibitem{garcia2000influence}
J.~Garc{\'i}a-Otero, M.~Porto, J.~Rivas, and A.~Bunde.
\newblock Influence of dipolar interaction on magnetic properties of ultrafine
  ferromagnetic particles.
\newblock {\em Physical Review Letters}, 84(1):167--170, 2000.

\bibitem{iglesias2004magnetic}
{\`O}.~Iglesias and A.~Labarta.
\newblock Magnetic relaxation in terms of microscopic energy barriers in a
  model of dipolar interacting nanoparticles.
\newblock {\em Physical Review B}, 70(14):144401, 2004.

\bibitem{kechrakos2005monte}
D.~Kechrakos and K.N. Trohidou.
\newblock Monte carlo study of the magnetic behavior of self-assembled
  nanoparticles.
\newblock {\em Journal of Magnetism and Magnetic Materials}, 295:177--179,
  2005.

\bibitem{weis2003simulation}
J.J. Weis.
\newblock Simulation of quasi-two-dimensional dipolar systems.
\newblock {\em Journal of Physics: Condensed Matter}, 15:S1471, 2003.

\bibitem{bujan2008influence}
M.C. Buj{\'a}n-N{\'u}{\~n}ez, N.~Fontai{\~n}a-Troiti{\~n}o,
  C.~V{\'a}zquez-V{\'a}zquez, M.A. L{\'o}pez-Quintela, Y.~Pi{\~n}eiro,
  D.~Serantes, D.~Baldomir, and J.~Rivas.
\newblock Influence of the nanoparticle size on the blocking temperature of
  interacting systems: Monte carlo simulations.
\newblock {\em Journal of Non-Crystalline Solids}, 354(47-51):5222--5223, 2008.

\bibitem{hoppe2008effect}
C.E. Hoppe, F.~Rivadulla, M.~Arturo Lo?pez-Quintela, M.~Carmen~Buja?n,
  J.~Rivas, D.~Serantes, and D.~Baldomir.
\newblock Effect of submicrometer clustering on the magnetic properties of
  free-standing superparamagnetic nanocomposites.
\newblock {\em The Journal of Physical Chemistry C}, 112(34):13099--13104,
  2008.

\bibitem{lamba2004single}
S.~Lamba and S.~Annapoorni.
\newblock Single domain magnetic arrays: role of disorder and interactions.
\newblock {\em The European Physical Journal B - Condensed Matter and Complex
  Systems}, 39(1):19--25, 2004.

\bibitem{oono1988study}
Y.~Oono and S.~Puri.
\newblock Study of phase-separation dynamics by use of cell dynamical systems.
  i. modeling.
\newblock {\em Physical Review A}, 38(1):434--453, 1988.

\bibitem{puri1988study}
S.~Puri and Y.~Oono.
\newblock Study of phase-separation dynamics by use of cell dynamical systems.
  ii. two-dimensional demonstrations.
\newblock {\em Physical Review A}, 38(3):1542--1565, 1988.

\bibitem{paterno2010magnetic}
L.G. Paterno, M.A.G. Soler, F.J. Fonseca, J.P. Sinnecker, E.H.C.P. Sinnecker,
  E.C.D. Lima, S.N. Bao, M.A. Novak, and P.C. Morais.
\newblock Magnetic nanocomposites fabricated via the layer-by-layer approach.
\newblock {\em Journal of Nanoscience and Nanotechnology}, 10(4):2679--2685,
  2010.

\bibitem{paterno2009layer}
L.G. Paterno, M.A.G. Soler, F.J. Fonseca, J.P. Sinnecker, E.H.C.P. Sinnecker,
  E.C.D. Lima, M.A. Novak, and P.C. Morais.
\newblock Layer-by-layer assembly of bifunctional nanofilms:
  Surface-functionalized maghemite hosted in polyaniline.
\newblock {\em The Journal of Physical Chemistry C}, 113(13):5087--5095, 2009.

\bibitem{soler2012assembly}
M.A.G. Soler, L.G. Paterno, J.P. Sinnecker, J.G. Wen, E.H.C.P. Sinnecker, R.F.
  Neumann, M.~Bahiana, M.A. Novak, and P.C. Morais.
\newblock Assembly of $\gamma-\mbox{Fe}_2\mbox{O}_3$/polyaniline nanofilms with
  tuned dipolar interaction.
\newblock {\em Journal of Nanoparticle Research}, 14(3):1--10, 2012.

\bibitem{morais2006preparation}
P.C. Morais, R.L. Santos, A.C.M. Pimenta, R.B. Azevedo, and E.C.D. Lima.
\newblock Preparation and characterization of ultra-stable biocompatible
  magnetic fluids using citrate-coated cobalt ferrite nanoparticles.
\newblock {\em Thin Solid Films}, 515(1):266--270, 2006.

\bibitem{sherman1985electronic}
D.M. Sherman and T.D. Waite.
\newblock Electronic spectra of $\mbox{Fe}^{+3}$ oxides and oxide hydroxides in
  the near ir to near uv.
\newblock {\em American Mineralogist}, 70(11-12):1262--1269, 1985.

\bibitem{bahiana1990cell}
M.~Bahiana and Y.~Oono.
\newblock Cell dynamical system approach to block copolymers.
\newblock {\em Physical Review A}, 41(12):6763, 1990.

\bibitem{shinozaki1993spinodal}
A.~Shinozaki and Y.~Oono.
\newblock Spinodal decomposition in 3-space.
\newblock {\em Physical Review E}, 48(4):2622, 1993.

\bibitem{hoshen1976percolation}
J.~Hoshen and R.~Kopelman.
\newblock Percolation and cluster distribution. i. cluster multiple labeling
  technique and critical concentration algorithm.
\newblock {\em Physical Review B}, 14(8):3438, 1976.

\bibitem{cullity2011introduction}
B.D. Cullity and C.D. Graham.
\newblock {\em Introduction to magnetic materials}.
\newblock Wiley-IEEE Press, 2011.

\bibitem{dutta2004magnetic}
P.~Dutta, A.~Manivannan, M.S. Seehra, N.~Shah, and G.P. Huffman.
\newblock Magnetic properties of nearly defect-free maghemite nanocrystals.
\newblock {\em Physical Review B}, 70(17):174428, 2004.

\bibitem{hendriksen1994ultrafine}
P.V. Hendriksen, F.~Bodker, S.~Linderoth, S.~Wells, and S.~Morup.
\newblock Ultrafine maghemite particles. i. studies of induced magnetic
  texture.
\newblock {\em Journal of Physics: Condensed Matter}, 6:3081, 1994.

\bibitem{hergt2004maghemite}
R.~Hergt, R.~Hiergeist, I.~Hilger, W.A. Kaiser, Y.~Lapatnikov, S.~Margel, and
  U.~Richter.
\newblock Maghemite nanoparticles with very high ac-losses for application in
  rf-magnetic hyperthermia.
\newblock {\em Journal of Magnetism and Magnetic Materials}, 270(3):345--357,
  2004.

\bibitem{jonsson1995aging}
T.~Jonsson, J.~Mattsson, C.~Djurberg, F.A. Khan, P.~Nordblad, and P.~Svedlindh.
\newblock Aging in a magnetic particle system.
\newblock {\em Physical Review Letters}, 75(22):4138--4141, 1995.

\bibitem{martinez1998low}
B.~Mart{\'i}nez, X.~Obradors, L.~Balcells, A.~Rouanet, and C.~Monty.
\newblock Low temperature surface spin-glass transition in
  $\gamma-\mbox{Fe}_2\mbox{O}_3$ nanoparticles.
\newblock {\em Physical Review Letters}, 80(1):181, 1998.

\bibitem{martinez1996magnetic}
B.~Martinez, A.~Roig, X.~Obradors, E.~Molins, A.~Rouanet, and C.~Monty.
\newblock Magnetic properties of $\gamma-\mbox{Fe}_2\mbox{O}_3$ nanoparticles
  obtained by vaporization condensation in a solar furnace.
\newblock {\em Journal of Applied Physics}, 79(5):2580--2586, 1996.

\bibitem{shendruk2007effect}
T.N. Shendruk, R.D. Desautels, B.W. Southern, and J.~Van~Lierop.
\newblock The effect of surface spin disorder on the magnetism of
  $\gamma-\mbox{Fe}_2\mbox{O}_3$ nanoparticle dispersions.
\newblock {\em Nanotechnology}, 18:455704, 2007.

\bibitem{vassiliou1993magnetic}
J.K. Vassiliou, V.~Mehrotra, M.W. Russell, E.P. Giannelis, R.D. McMichael, R.D.
  Shull, and R.F. Ziolo.
\newblock Magnetic and optical properties of $\gamma-\mbox{Fe}_2\mbox{O}_3$
  nanocrystals.
\newblock {\em Journal of Applied Physics}, 73(10):5109--5116, 1993.

\bibitem{nowak2000monte}
U.~Nowak, R.W. Chantrell, and E.C. Kennedy.
\newblock Monte carlo simulation with time step quantification in terms of
  langevin dynamics.
\newblock {\em Physical Review Letters}, 84(1):163, 2000.

\bibitem{papusoi1999particle}
C.~Papusoi.
\newblock The particle interaction effects in the field-cooled and
  zero-field-cooled magnetization processes.
\newblock {\em Journal of Magnetism and Magnetic Materials}, 195(3):708--732,
  1999.

\end{thebibliography}
\end{document}